\documentclass{iopjournal}
\usepackage{latexsym}
\usepackage{amssymb}
\begin{document}

\articletype{Paper} 

\title{Emergence of oscillatory states of self-propelled colloids under optical confinement}

\author{Farshad Darabi 
and Juan Ruben Gomez-Solano$^{*}$\orcid{0000-0003-2475-1151}}

\affil{Instituto de Física, Universidad Nacional Autónoma de México, Ciudad de México, Código Postal 04510, Mexico.}

\affil{$^*$Author to whom any correspondence should be addressed.}

\email{r\_gomez@fisica.unam.mx}


\begin{abstract}
We investigate experimentally the single-particle motion in water of silica colloidal beads half-coated with carbon under the action of a converging laser beam. The beads are self-propelled in this medium by means of self-thermophoresis resulting from local heating as a result of light absorption by their carbon cap. Within a certain laser power range, we find that these particles exhibit a quasi-two-dimensional active motion near a solid surface with stochastic rotational reversals when propelling themselves away from the region of maximum intensity, which leads to a stable trapping with oscillatory-like behavior inside the illuminated region. The orientational autocorrelation function of this type of confined active motion displays damped oscillations whose characteristic frequency increases with increasing propulsion speed, thus resulting in four regimes of translational motion depending on the observation time scale: thermal diffusion, ballistic motion, oscillatory behavior, and confinement. Our experimental findings are well described by a minimal phenomenological model that includes the nonlinear effect of a torque that reorients the particle toward the center of the optical confinement, which in combination with rotational diffusion gives rise to the observed orientational changes that allow their oscillatory trapping inside the light field. We also show that a similar active trapping mechanism emerges in the
case of Janus colloidal rods, even though the periodicity is hindered by their three-dimensional rotation in the laser beam.
\end{abstract}

\section{Introduction}\label{sect:intro}
Active colloidal systems, such as motile microorganisms
and artificial microswimmers, are a type of soft matter of utmost importance in several fields of fundamental and applied science because of their capability to autonomously convert the energy
harvested from their environments into persistent motion \cite{ramaswamy2010,elgeti2015,bechinger2016}.
Therefore, they can exhibit non-equilibrium collective phenomena without counterpart at thermal equilibrium, such as swarming \cite{thutupalli2011,peruani2012,bricard2013}, dynamic clustering \cite{theurkauff2012,buttinoni2013,cates2015}, rectification \cite{wan2008,dileonardo2010,koumakis2019}, active turbulence \cite{dunkel2013,nishiguchi2015,creppy2015,luo2025}
, etc. Even at the single-particle level, active colloids display complex behavior when self-propelling in fluids, like tactic responses to external gradients \cite{hong2007,palacci2015,gomezsolano2017,yu2019}, guidance by solid walls \cite{berke2008,das2015,simmchen2016}, and intricate swimming trajectories due their shape \cite{kuemmel2013,shemi2018,anchutkin2024} or when moving in viscoelastic media \cite{gomezsolano2016,narinder2018,taveravazquez2024,saito2025}, all of which result from their anisotropic coupling with their surroundings.

Of special significance is the response of active colloids to externally applied fields, as its understanding is crucial for their manipulation in fluid environments by, e.g. acoustic \cite{takatori2016} and magnetic forces \cite{pierce2017}. In particular, Janus colloids, which are synthetically made of a dielectric material half-coated by a thin cap with distinct physical properties \cite{walther2008,wang2020}, represent an ideal experimental system that can help elucidate various non-equilibrium features of active particles in confining potentials \cite{solon2015,jahanshahi2017,das2018,basu2019,malakar2020,santra2021,nakul2023,caprini2023,alexandre2024,adersh2024,arredondo2024}. If such particles are exposed to light fields like those created by tightly focused laser beams, optical forces and torques can be exerted on them \cite{gieseler2021}, while their self-propulsion through fluids can be induced by, e.g. self-thermophoresis due to light absorption by their coating \cite{jiang2010,bickel2013,bregulla2015}, self-electrophoresis and self-diffusiophoresis chemically catalyzed by the cap surface \cite{howse2007,brown2014,raman2023}. When the torques exerted by optical tweezers are negligible, Janus particles must behave almost perfectly as active Brownian particles in a harmonic potential, whose swimming direction is solely governed by rotational diffusion. Indeed, this has been verified in recent experiments using silica-capped polymethyl methacrylate particles with a quasi-homogeneous refractive index activated by electro-osmotic flows \cite{buttinoni2022}, and self-diffusiophoretic platinum-capped silica particles \cite{halder2025}. In such experiments, non-equilibrium bimodal distributions of the particle positions predicted by theoretical models are observed, which result from the interplay between the viscous relaxation time and the persistence time of the self-propulsive force determined by the inverse of the rotational diffusion coefficient. Under such conditions, the self-diffusiophoretic forces that lead to the active motion of catalytic Janus colloids can be measured directly using optical tweezers \cite{raj2025}.

Nevertheless, it is important to point out that in most experimental situations of interest, Janus particles possess highly anisotropic properties in the presence of light. Therefore, their dynamical behavior in confining light fields is not simply dictated by rotational diffusion but is strongly affected by orientation-dependent torques of optical and thermal origin. For instance, gold-capped dielectric beads generally exhibit optical trapping with orbital motion in a tightly focused laser beam at moderate intensity \cite{merkt2006,liu2015,zong2015}, where persistent rotations are a consequence of the optical torques mainly originating from the high reflectivity of their metallic cap. This type of orbital motion in optical tweezers has even been exploited to realize microengines based on self-propelled Janus particles with light-absorbing caps, whose rotation can be controlled by temperature \cite{schmidt2018} and light polarization \cite{bronte2023}. Furthermore, in extended light patterns, self-thermophoretic Janus particles can display more intricate behavior, such as round-trip motion in static line optical tweezers \cite{liu2016}, trochoidal trajectories in a diverging laser beam \cite{moyses2016}, and moon-like revolution and rotation of in an annular optical trap \cite{liu2024}.
In the majority of the aforementioned systems, the direction of rotation along the particle trajectories remains unchanged because of the dominance of persistent optical torques over thermal fluctuations.

In this paper, we report a novel type of oscillatory-like motion
of active Janus colloids in a confining optical potential. To this
end, we conducted experiments of silica beads half-coated with
carbon that actively move by self-thermophoresis in water under
the action of a converging laser beam. We find that these particles
can be dynamically trapped by this light field, displaying a quasi-two-dimensional oscillatory
active motion, which results from the interplay between their
self-propulsion and the optical forces and torques exerted by the
laser beam. Our results are well described by a phenomenological
model that includes a torque that reorients an active Brownian
particle toward the center of a harmonic potential. In addition,
we show that a similar back-and-forth motion also occurs for
rod-shaped Janus particles actively moving in the same type of laser beam.

\section{Experimental methods}\label{sect:exp}

\begin{figure}[h]
\centering
\includegraphics[width=0.85\columnwidth]{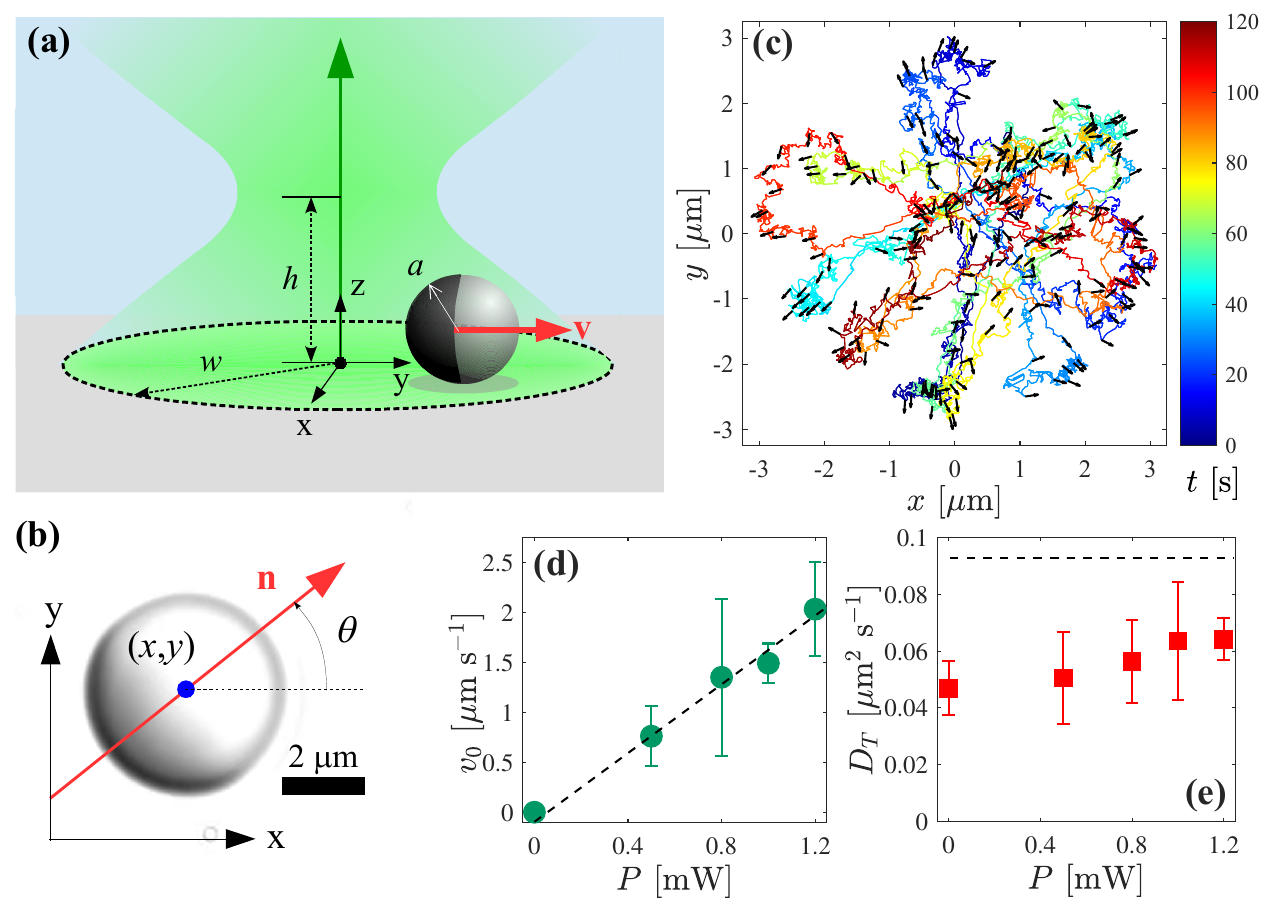}
  \caption{(a) Sketch of a spherical Janus bead (radius $a = 2.32$~µm) self-propelling in water with velocity {\bf{v}} due to the action of a laser beam (wavelength $\lambda = 532$~nm), which propagates in the $+z$ direction and is tightly focused inside the sample cell at a distance of $h = 7$~µm from its lower inner surface, thus illuminating a circular area of radius $w \approx 10$~µm. The coordinate system $xyz$ used to characterize the particle motion is also outlined. (b) Snapshot of a silica bead half-coated with a 100~nm thick carbon layer (dark side). The blue spot represents the coordinates ${\bf{r}}= (x,y)$ of its center of mass, while the red arrow depicts the projection onto the $xy$ plane of its orientation $\mathbf{n}$, which defines the angle $\theta$. (c) Example of a trajectory segment of a Janus bead moving actively in the converging laser beam at $P = 0.5$~mW. The color map portrays the time elapsed over 2 minutes, $0 \le t \le 120$~s. The arrows indicate the instantaneous particle orientation every 0.5 seconds. (d) Dependence of the characteristic propulsion speed of the Janus particles on the applied laser power. The dashed line is a linear fit of the experimental data. (e) Translational diffusion coefficient of the Janus particles at short time scales as a function of the laser power. The horizontal dashed line represents the value estimated in the bulk by means of the Stokes-Einstein relation.}
  \label{fig:1}
\end{figure}

In our experiments, we used Janus colloidal particles made of silica microspheres (diameter $2a=4.64$~µm) half-coated with a carbon layer (thickness 100~nm) that was deposited by sputtering (Ernst F. Fullam, Inc.), resulting in a hemispherical cap. These Janus beads were dispersed in ultrapure water (resistivity 18.2~M$\Omega$~cm at 25$^{\circ}$C) at a concentration of less than 1 particle per nl of solution. The diluted colloidal dispersion was confined in a sample cell consisting of a microscope glass slide attached to a coverslip by double-sided adhesive tape (thickness $\sim 100\,\mu$m) and sealed with epoxy glue to prevent leakage and evaporation.

A green laser beam with Gaussian profile and wavelength of 532~nm (Opus 532, Laser Quantum) was used to create a trapping potential for a single Janus particle and also to induce its self-propulsion in water. 
To do this, the laser beam was focused by an oil immersion objective (Olympus Plan Fluorite 100$\times$, NA = 1.3) inside the sample cell at a distance $h = 7$~µm from the inner surface of the coverslip, as sketched in figure \ref{fig:1}(a). Because of the effect of sedimentation, the Janus particles were not trapped close to the beam focus by the vertical component of the gradient forces as in conventional optical tweezers experiments, but remained close to the lower glass surface. Under such conditions, the radiation pressure exerted by the converging beam provides the particle with an optical force whose component parallel to the $xy$ plane is radially directed towards the $z$ axis ($x=0, y=0)$, i.e. perpendicular to the beam propagation direction, effectively resulting in a confining potential inside an illuminated circular region of radius $w \approx 10$~µm. In addition, the absorption of light by the carbon cap generates an anisotropic temperature gradient around the particle that is dependent on the incident power, which in turn leads to
its self-propulsion due to thermophoretically induced flows \cite{bickel2013}.   
All experiments were carried out at room temperature ($T = 22 \pm 0.2^\circ$C), at which the dynamic viscosity of the bulk water within the sample cell is $\eta = 0.953 \pm 0.004$~mPa~s.

The translational and rotational motion of a single particle trapped in the converging laser beam was recorded using a CMOS camera (Basler acA800-510um) at a sampling frequency of 100 frames per second. Videos of distinct independent particles at varying laser powers were recorded in order to investigate their resulting dynamics under diverse confinement and self-propulsion conditions.
From the recorded videos, at each frame we detected the projection onto the $xy$ plane of the position of the particle, ${\bf{r}}=(x,y)$, by means of standard particle-tracking algorithms with a spatial resolution of 5~nm. In addition, without requiring additional image enhancement, the contrast between the dark carbon cap and the brighter silica hemisphere enabled accurate tracking of the projection $\mathbf{n} = (\cos \theta,\sin \theta)$ onto the $xy$ plane of the instantaneous particle orientation, defined as the unitary vector that points from the carbon to the silica hemisphere, as illustrated in figure \ref{fig:1}(b). Each video recording lasted $t_{tot}\approx 15-30$~minutes, thus providing sufficient data to analyze the particle dynamics across multiple timescales. An example of a section of a typical particle trajectory with the corresponding orientation vector over 2 minutes is depicted in figure \ref{fig:1}(c), which clearly illustrates the confined active motion induced by the converging laser beam within a region of a few microns. It should be noticed that such active trajectories under optical confinement are markedly distinct from the Brownian motion of uncapped silica particles at thermal equilibrium in a harmonic optical potential, which follows almost perfectly an Ornstein-Uhlenbeck process~\cite{gieseler2021}. Instead, the carbon-capped particles in the same laser beam perform segments of directed motion as they traverse the central part of the illuminated area, whose persistence is interrupted by pronounced directional changes occurring when they move away from the beam axis.   

To change the characteristic propulsion speed of the Janus particles as well as the strength of the optical confinement,
the laser power was adjusted between 0 and 1.2~mW. For each power, five independent trials were conducted with different Janus beads. 
We observed that trapping was not possible at powers below 0.2~mW because Janus particles propel themselves far from the central part of the light field to regions where no active motion is induced. We also checked that uncapped silica beads could not be trapped either, thus verifying that optical forces become negligible at such low values of the laser power. Moreover, at laser powers above 1.2 mW, we noticed that the radiation pressure in the $+z$ direction is sufficiently strong to surpass the particle weight, thus pushing it away from the confining region of the potential. Therefore, stable optical trapping of the carbon-capped particles is not easily achievable at high laser power. Within the power range in which the Janus colloidal beads exhibit persistent confinement ($0.2\,\mathrm{mW} < P \le 1.2\,\mathrm{mW}$), we determined their characteristic propulsion speed, $v_0$, from the velocity autocorrelation function of the particle (see Appendix \ref{app:1}), thereby finding a linear increase with the laser power, $P$, as demonstrated in figure \ref{fig:1}(d). Note that this linear increase is in agreement with the dependence of the propulsion speed on the applied laser power expected for self-thermophoretic Janus colloids with a light-absorbing cap moving in water. In such a case, the temperature difference across the particle, which is in turn proportional to the propulsion speed, depends linearly on the power absorbed by the cap~\cite{jiang2010,bickel2013}.
In addition, from the velocity autocorrelation function, we could also measure the short-time translational diffusion coefficient of the particles, $D_T$, for the different laser powers, which are plotted in figure \ref{fig:1}(e). Note that these values are all smaller than that estimated in the bulk using the Stokes-Einstein relation ($\frac{k_B T} {6\pi \eta a} = 0.0929$~µm$
^2\,\mathrm{s}
^{-1}$) due to the proximity of the particles to the bottom wall of the sample cell, which increases the effective friction on them. Furthermore, the measured values of $D_T$ slightly increase with higher laser power because of the temperature dependence of the water viscosity. From these data, we estimate that the local temperature of water around the particle heated by light absorption by the carbon cap increases at most $11$~K at $P = 1.2$~mW relative to the bulk temperature.

\section{Results and discussion}\label{sect:res}

\begin{figure}[h]
\centering
\includegraphics[width=0.9\columnwidth]{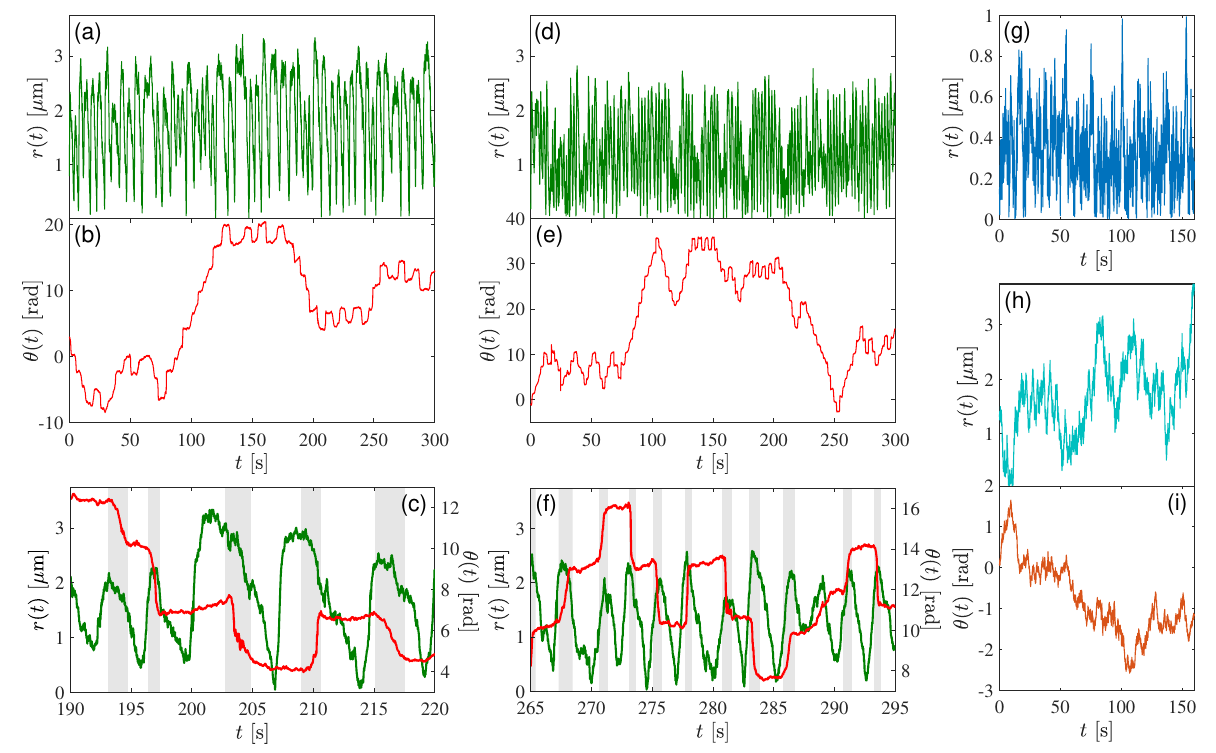}
  \caption{[(a)-(c)] Stochastic time evolution over 300~s of the radial position of a Janus particle relative to the beam axis (a) and its orientation angle (b) under the action of the converging laser beam at $P = 0.5$~mW. An expanded view of both coordinates [$r(t)$ in green and $\theta(t)$ in red] during 30~s is shown in (c). [(d)-(f)] Stochastic motion over 300~s of the radial position of the same Janus particle (d) and its orientation angle (e) subject to the converging beam at $P = 1.0$~mW. An enlarged view of both coordinates [$r(t)$ in green and $\theta(t)$ in red] during 30~s is shown in (f). The shaded areas in (c) and (f) depict sudden orientational changes of $|\delta \theta| \approx \pi$~rad occurring when the particle reaches comparatively large radial distances. (g) Stochastic evolution during 150~s of the radial distance of an uncoated silica bead in the laser beam at $P=0.5$~mW. [(h)-(i)] Free motion over 150~s of the radial position of a Janus bead (h) and its orientation angle (i) at $P = 0$.}
  \label{fig:2}
\end{figure}

\subsection{Translational and rotational motion}\label{subsect:rottrans}

In figures \ref{fig:2}(a)-(f) we represent examples of the stochastic evolution with time $t$ of the Janus particle's radial position relative to the center of the beam axis located at $(x=0,y=0)$, $r(t) = |{\bf{r}}(t)|= \sqrt{x(t)^2 + y(t)^2}$, and the angle $\theta(t)$ defined by the particle orientation for two distinct laser powers ($P = 0.5$~mW, 1.0~mW). For comparison, in figure \ref{fig:2}(g) we plot the radial position of an uncoated silica bead exposed to the same beam at $P = 0.5$~mW, which undergoes diffusion confined to a harmonic potential. Moreover, in figures \ref{fig:2}(h) and \ref{fig:2}(i) we trace the radial position  and orientation angle of a free Janus particle at $P=0$, whose stochastic dynamics are dictated only by translational and rotational diffusion, respectively. As can be seen in figures \ref{fig:2}(a)-(f), a non-equilibrium oscillatory motion of the Janus beads becomes manifest in the presence of the converging laser, which is absent in the equilibrium cases of uncapped particles in an optical potential and freely diffusing carbon-coated particles shown in figures \ref{fig:2}(g)-(i).
Interestingly, the active oscillatory dynamics of the carbon-coated particles in the laser beam also differs substantially from that expected for an active Brownian particle subject to rotational diffusion in a harmonic trap \cite{solon2015,das2018,basu2019,malakar2020,nakul2023} or from the orbiting motion of metal-capped beads observed in optical tweezers experiments \cite{merkt2006,liu2015,zong2015,bronte2023}. In such instances, the active particles rarely visit the central part of the region illuminated by the laser spot but remain on average at a non-zero radial distance from the beam axis where optical and propelling forces balance each other. In contrast, the carbon-capped particles studied here can travel back and forth within the region illuminated by the laser, as shown in figure \ref{fig:1}(c), thus passing frequently through its central part, see also figures \ref{fig:2}(a), \ref{fig:2}(c), \ref{fig:2}(d), and \ref{fig:2}(f). Furthermore, the comparison between the radial position $r(t)$ and the angle $\theta(t)$ in figures \ref{fig:2}(c) and \ref{fig:2}(f) demonstrates that while the angle does not change appreciably when the particle moves toward or away from the beam axis (unshaded areas), it performs a rotation of approximately $\pi$~radians when the particle reaches comparatively large radial distances (shaded areas), thus almost fully reversing its direction of motion. Interestingly, such orientational reversals can occur randomly clockwise ($\delta \theta \approx -\pi$) or anticlockwise ($\delta \theta \approx +\pi$) along the same trajectory. Therefore, a direct consequence of this is that $\theta(t)$ becomes fully randomized at sufficiently long time-scales, as evidenced in figures \ref{fig:2}(b) and \ref{fig:2}(e). Note that this behavior of $\theta$ is different from that observed for metal-capped particles under optical confinement, where rotational motion in a single direction occurs commonly in the absence of external torques~\cite{merkt2006,liu2015,zong2015,schmidt2018,bronte2023,liu2016}. Additionally, we observed that for all the explored laser powers at which the particles can be stably trapped, their orientation vector remains almost parallel to the $xy$ plane, and therefore we can assume that their rotational dynamics takes place mostly in two dimensions, then being well characterized by the single angle $\theta$. Thus, the joint effect of the thermophoretic self-propulsion, the stochastic orientational reversals, and the confinement created by the laser beam suggests the emergence of oscillatory-like behavior in the active motion of these Janus colloidal beads in the localized two-dimensional area defined by the converging beam.

\begin{figure}[h]
\centering
\includegraphics[width=0.9\columnwidth]{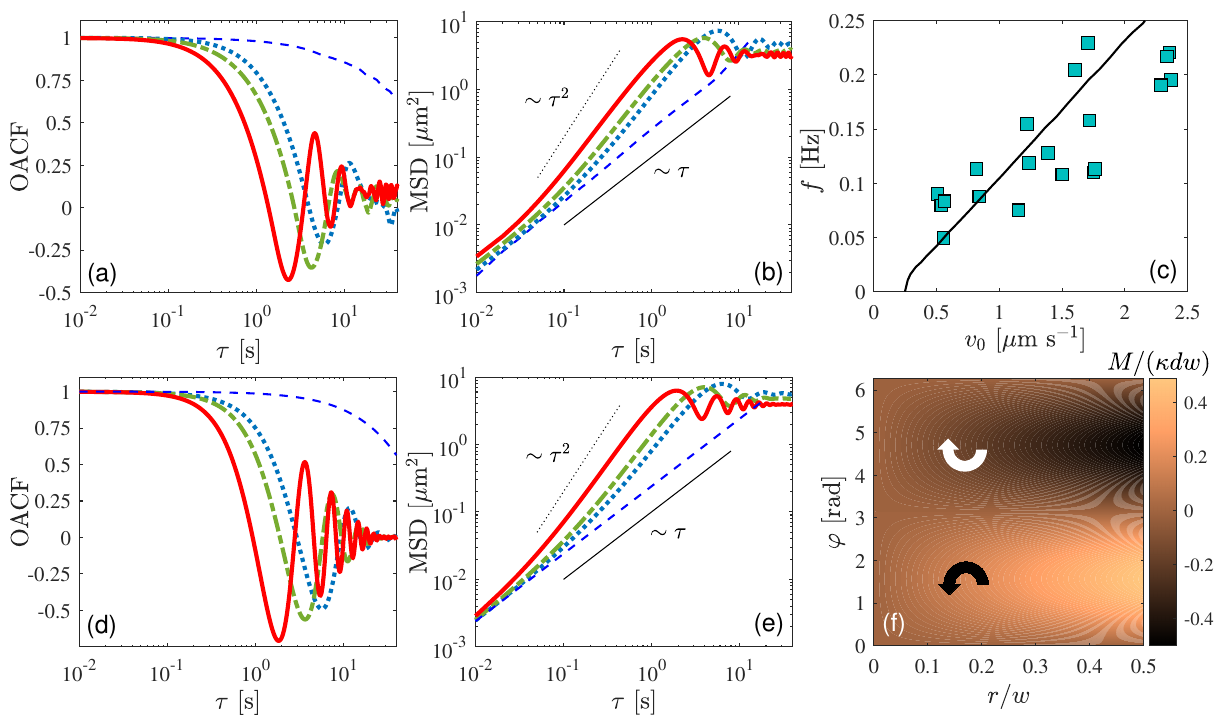}
  \caption{(a) Experimental orientation autocorrelation function of the Janus beads defined by equation (\ref{eq:aacf}) for three representative propulsion speeds achieved in the converging laser beam: $v_0= 0.840$~µm~s$^{-1}$ (dotted line), 1.238~µm~s$^{-1}$ (dotted-dashed line), 2.334~µm~s$^{-1}$ (solid line). The dashed line represents the exponentially decaying behavior with relaxation time $\tau_r \approx 70$~s of the OACF experimentally measured for a Janus bead in the absence of the laser beam ($v_0=0$). (b) Experimental mean-square displacement of the Janus particle position for the same propulsion speeds shown in figure \ref{fig:3}(a) represented with the same line style. Thin solid and dotted lines depict diffusive and ballistic behavior, respectively. (c) Dependence of the characteristic oscillation frequency of active Janus particles under the optical confinement on their propulsion speed determined from the corresponding experimental OACF curve ($\Box$). Each data point results from the analysis of a single trajectory of an independent particle moving at a fixed laser power. The solid line is a numerical curve calculated using the stochastic model described by equation (\ref{eq:xy}) and (\ref{eq:theta}). [(d)-(e)] Numerical results of the simulation of the stochastic dynamics described by the model of equation (\ref{eq:xy}) and (\ref{eq:theta}): (d) orientation autocorrelation function, and (e) positional mean-square displacement. The values of the propulsion speed used in the simulation are the same as those determined in the experiments represented in figures \ref{fig:3}(a) and \ref{fig:3}(b), and are also depicted with the same line style. (f) Colormap representing the magnitude $M$ of the torque (normalized by $\kappa d w$) relative to the direction $+z$ that is included in the rotational part of the model, see equation (\ref{eq:theta}), as a function of the radial position $r$ of the particle (normalized by $w$) and the angle $\varphi$ between the particle orientation, ${\bf{n}}$, and minus its position, $-{\bf{r}}$. The arrows indicate the sense of rotation of ${\bf{n}}$ in the $xy$ plane induced by the torque depending on $\varphi$: counterclockwise ($0\le \varphi < \pi$), and clockwise ($\pi < \varphi \le 2\pi$).}
  \label{fig:3}
\end{figure}

To quantify this nontrivial dynamical behavior, we calculate the orientation autocorrelation function (OACF) measured at two times, $t$ and $t + \tau$
\begin{equation}\label{eq:aacf}
\langle \mathbf{n}(t+\tau) \cdot \mathbf{n}(t)\rangle_t = \langle \cos [ \theta(t+\tau) - \theta(t)] \rangle_t,
\end{equation}
like the curves illustrated in figure \ref{fig:3}(a) for a particle actively moving in this light field at varying laser powers $P > 0.2$~mW resulting in three distinct values of $v_0$. In equation (\ref{eq:aacf}), $\langle \ldots \rangle_t$ denotes an average over the initial time $t$ along a complete trajectory of a specific particle with duration
$t_{tot}$, i.e. $0 \le t \le t_{tot} - \tau$. We find that in all cases the OACF exhibits an oscillatory decay as a function of the time lag $\tau$ with a well-defined period $\mathcal{T}$ that depends on $v_0$, where the oscillations are damped at sufficiently large timescales. 
This is in stark contrast to the rotational motion in the absence of the converging beam, whose OACF corresponds to a decaying exponential function with a single relaxation time $\tau_R \approx 70$~s approximately determined by the inverse of the rotational diffusion coefficient $D_R = k_B T/(8\pi \eta a^3) \approx 0.014\,\mathrm{rad}\,\mathrm{s}^{-1}$, as shown by the dashed line in figure \ref{fig:3}(a).

The oscillatory behavior of the OACF of the carbon-coated particles in the confining optical potential has important consequences for the dependence on the time difference $\tau$ of the mean-square displacement (MSD) of their two-dimensional position ${\bf{r}} = (x,y)$
\begin{equation}\label{eq:MSD}
    \langle |\Delta \mathbf{r}(\tau)|^2 \rangle = \langle [x(t+\tau)-x(t)]^2 + [y(t+\tau) - y(t)]^2 \rangle_t, 
\end{equation}
as illustrated in figure \ref{fig:3}(b) for the same three different values of the propulsion speed analyzed in figure \ref{fig:3}(a). Interestingly, the MSD discloses four distinctive regimes of the translational motion of the Janus particles under these experimental conditions. First, on very short timescales, the particle displays diffusive behavior, i.e. $\langle |\Delta \mathbf{r}(\tau)|^2 \rangle \sim \tau$, due to the thermal collisions of the surrounding water molecules, where the corresponding diffusion coefficient estimated through the velocity autocorrelation function was already shown in figure \ref{fig:1}(e). For comparison, the MSD of a purely diffusive Janus particle in the absence of the laser beam is also displayed in figure \ref{fig:3}(b). Then, at intermediate timescales, the motion becomes ballistic, i.e. $\langle |\Delta \mathbf{r}(\tau)|^2 \rangle \sim \tau^2$, with a well-defined speed whose values are close to those determined by means of the velocity autocorrelation function. This regime mirrors the behavior of particle trajectory segments with durations over which directional persistence occurs, i.e. without significant variations of the particle orientation that can originate from rotational diffusion or optical torques. Moreover, with increasing $\tau$ an oscillatory behavior of the MSD driven by the combined action of self-propulsion and optical forces and torques exerted by the laser beam becomes evident. The period $\mathcal{T}$ of this oscillatory regime of the MSD coincides with that related to the OACF, i.e. with the characteristic timescale of orientational reversals when the particles propel themselves away from the beam center. Finally, at sufficiently long timescales, the oscillations are gradually damped, which results in a saturation of the MSD to a constant time-independent value, thus confirming the dynamical confinement of the Janus particles inside the region illuminated by the laser.

From the temporal separation between consecutive maxima of the experimental OACF curves, or equivalently from that between consecutive minima of the respective MSD, we determine the period $\mathcal{T}$, whose inverse $f = 1/\mathcal{T}$ corresponds to the oscillation frequency  of the active motion in the converging beam. The dependence of $f$ on the propulsion speed $v_0$ is plotted as squares in figure \ref{fig:3}(c), where each data point is the result of the analysis  over the complete trajectory of a specific particle self-propelling at a fixed laser power, giving rise to values of $v_0$ falling within the interval $0.5\,\mu\mathrm{m} \, \mathrm{s}^{-1} \lesssim v_0 \lesssim 2.5\,\mu\mathrm{m} \, \mathrm{s}^{-1}$. The experimental data of figure \ref{fig:3}(c) show that, on average, $f$ increases monotonically with $v_0$. 
Data dispersion is likely due to small variations ($5\%$) in particle size around the nominal diameter of $4.64\,\mu$m and in carbon-coating thickness, but a monotonically increasing trend of $f$ as a function of $v_0$ can be clearly detected.
Note that at sufficiently small propulsion speeds, which correspond to laser powers below the minimum threshold value for optical trapping ($P_0 = 0.2$~mW), the oscillatory behavior does not occur even if the particles self-propel across the region illuminated by the laser spot because they move to regions where the laser intensity decays to zero, thus becoming non-active. This confirms that the self-sustained oscillations that we observe in the experiments emerge from the interplay between the activity of the Janus particle and the confining optical forces and torques acting on it. 

\subsection{Model}\label{subsect:mod}

To explain the experimental results that were previously presented, we consider a minimal phenomenological model for the overdamped motion of the Janus particles in the converging laser beam. First, we note that for all the explored laser powers at which stable trapping is achieved, the particles move in a localized region of radius of at most $\approx 3$~µm. Within this region, owing to the Gaussian profile of the laser beam whose width is $w \approx 10~\mu$m, its intensity decreases at most 5~\% relative to the maximum intensity at $x=y=0$. Therefore, we can assume that for a fixed value of the laser power, the propulsion speed of the active particles does not depend on their radial position and is given by the characteristic speed $v_0$ plotted in figure \ref{fig:1}(d), i.e.
\begin{equation}\label{eq:vel}
    v_0 = \alpha P - b,
\end{equation}
where $\alpha = 1.723$~µm~mW$^{-1}$ and $b = 0.096$~µm~s$^{-1}$. As a consequence, the self-propelling velocity at time $t$, ${\bf{v}}(t)$, can be considered to behave as in a quasi-uniform environment, where it must point parallel to the instantaneous particle orientation, i.e. ${\bf{v}}(t) = v_0 {\bf{n}}(t)$. Second, since the optical force exerted by the laser at time $t$, ${\bf{F}}_{\rm{opt}}(t)$, leads to a confined active motion of the Janus particle, as a first approximation it can be supposed to behave as a linear restoring force, i.e. ${\bf{F}}_{\rm{opt}}(t) = -\kappa \bf{r}(t)$, similar to the effect of a focused laser beam on the motion of platinum-silica beads \cite{halder2025}. According to the experimental observations, the trapping stiffness is related to the applied laser power as
\begin{equation}\label{eq:kappa}
    \kappa = \beta (P - P_0),
\end{equation}
where $P_0 = 0.2$~mW is the threshold power above which optical confinement can be achieved and $\beta$ is a constant parameter to be determined. Then, the trapping stiffness can be expressed in terms of the propulsion speed as the linear relation $\kappa = (\beta/\alpha) v_0 + \beta (b/\alpha -P_0)$. Moreover, by taking into account the friction force and thermal noise from the surrounding water molecules, the overdamped dynamics of ${\bf{r}}(t) = (x(t),y(t))$ can be modeled by the following stochastic differential equations
\begin{eqnarray}\label{eq:xy}
    \frac{d}{dt}x(t) & = &  v_0 \cos \theta(t) - \frac{\kappa}{\gamma_T} x(t) + \sqrt{2D_T} \zeta_x(t), \nonumber\\
    \frac{d}{dt}y(t) & = &  v_0 \sin \theta(t) - \frac{\kappa}{\gamma_T} y(t) + \sqrt{2D_T} \zeta_y(t).
\end{eqnarray}
In equation (\ref{eq:xy}), $D_T$ is the translational diffusion coefficient, which, for the sake of simplicity, can be assumed constant and  equal to the value $D_T = 0.050$~µm$
^2\,\mathrm{s}
^{-1}$ measured at $P=0$, see figure \ref{fig:1}(e); $\gamma_T = k_B T/D_T$ is the corresponding translational friction coefficient; $T = 295.15$~K is the bath temperature; $\zeta_x(t)$ and $\zeta_y(t)$ are Gaussian noises with zero mean, \textit{i.e.} $\langle \zeta_x(t) \rangle = \langle \zeta_y(t) \rangle = 0$, and correlations $\langle \zeta_x(t) \zeta_y(t') \rangle = 0$ and $\langle \zeta_x(t) \zeta_x(t') \rangle = \langle \zeta_y(t) \zeta_y(t') \rangle = \delta(t-t')$. 

Equation (\ref{eq:xy}) must be complemented by a stochastic equation of motion of the orientation angle $\theta(t)$. In such a case, we consider that as a consequence of the optical anisotropy created by the light-absorbing carbon cap, the optical force must exert a torque on the Janus particle.  The simplest form of this torque can be expressed as ${\bf{M}}(t) = {\bf{d}}(t) \times {\bf{F}}_{\rm{opt}}(t) \equiv M(t)\mathrm{\hat{z}}$, where ${\bf{d}}(t) = d {\bf{n}}(t)$ is an effective lever arm of magnitude $d \sim a$, whose direction is determined by the main anisotropy axis of the particle, i.e. ${\bf{n}}(t)$, and $\mathrm{\hat{z}}$ is the unit vector pointing in the direction $+z$. In this way, a stochastic differential equation that phenomenologically describes the dynamics of $\theta(t)$, is
\begin{equation}\label{eq:theta}
    \frac{d}{dt} \theta(t)  =  \frac{\kappa d}{\gamma_R} \left[ x(t) \sin \theta(t) - y(t) \cos \theta(t) \right] + \sqrt{2 D_R} \zeta_{\theta}(t),
\end{equation}
where $D_R$ is a rotational diffusion coefficient. In equation (\ref{eq:theta}), for the sake of simplicity, $\zeta_{\theta}(t)$ can also be assumed as a Gaussian delta-correlated noise with zero mean, \textit{i.e.} $\langle \zeta_{\theta}(t) \rangle = 0$, and $\langle \zeta_{\theta}(t) \zeta_{\theta}(t') \rangle = \delta(t-t')$.

We solved numerically the system of nonlinear stochastic differential equations (\ref{eq:xy}) and (\ref{eq:theta}) using the Euler–Maruyama method. We set the free parameters $\beta$ and $d$ to the values $\beta = 35\,\mathrm{fN}\,\mu\mathrm{m}^{-1} \, \mathrm{mW}^{-1}$ and $d = 0.7a$, which were selected to get the best agreement of the simulated results with the measured standard deviation of the radial position of the Janus particles and their characteristic oscillation frequencies, at all experimental values of the propulsion speed. From the simulated trajectories ${\bf{r}}(t)=(x(t),y(t))$ and $\theta(t)$ at each value of $v_0$, we calculate the orientation autocorrelation function and the mean-square displacement, as defined by equations (\ref{eq:aacf}) and (\ref{eq:MSD}), respectively. 

In figure \ref{fig:3}(d) we illustrate some numerical OACF curves obtained from simulated trajectories using the model for the same experimental values of the propulsion speed displayed in figure \ref{fig:3}(a). It is noteworthy that these numerical OACF curves resemble the experimental ones, where damped oscillations arise from the stochastic reversals of $\theta(t)$ with a frequency that increases with increasing $v_0$. Figure \ref{fig:3}(e) shows that the numerical MSD curves also reproduce the behavior observed in the experiments, namely, the four regimes of translational motion of carbon-coated particles that occur within the propulsion speed interval at which their optical trapping is possible. Moreover, using the same method as in the case of the experimental OACF and MSD curves, we calculate the dependence of the oscillation frequency on the propulsion speed, which is traced as a solid line in figure \ref{fig:3}(c). We confirm that the monotonic increase of $f$ with $v_0$ observed in the experimental data is well captured by the oscillatory dynamics resulting from the model described by equations (\ref{eq:xy}) and (\ref{eq:theta}). In particular, for $v_0 \gtrsim 0.5\,\mu\mathrm{m}\,\mathrm{s}^{-1}$ the model predicts a linear dependence of the oscillation frequency on the propulsion speed. 

We point out that the essential component of the particle dynamics that is necessary for the emergence of self-sustained active oscillations under optical confinement is the torque originating from the anisotropic refractive index of the Janus particle. In figure \ref{fig:3}(f) we plot as a colormap the values of the magnitude $M$ of the torque considered in the model (normalized by $\kappa dw$, with $\kappa > 0$), as a function of the radial distance of the particle, $r$, and the angle $\varphi$ between its orientation, ${\bf{n}}$, and minus its position vector $-{\bf{r}}$, the latter proportional to the restoring optical force ${\bf{F}}_{\rm{opt}} = -\kappa {\bf{r}}$. As can be seen, at any non-zero radial distance, the torque tends to align ${\bf{n}}$ parallel to $-{\bf{r}}$ ($\varphi = 0 = 2\pi$), the strongest torque occurring when $\varphi = \pi/2, 3\pi/2$, i.e. when the particle orientation is perpendicular to its position vector. Specifically, if $0 \le \varphi < \pi$, ${\bf{n}}$ has the tendency to rotate counterclockwise in the $xy$ plane ($M > 0$), while a clockwise rotation is induced if $\pi < \varphi \le 2\pi$ ($M<0$). Furthermore, for any $\varphi \neq \pi$ the torque becomes stronger when the particle is far from the center of the beam and decreases as it approaches it, vanishing at $r=0$. Note that even if an alignment of ${\bf{n}}$ antiparallel to $-{\bf{r}}$ results in a null torque when $\varphi = \pi$, such particle orientations are unstable to rotational perturbations. Therefore, these particular features of the torque considered in the model represent a plausible sensing mechanism that gives rise to the observed oscillatory self-trapping of carbon-capped particles without escaping by self-propulsion the confining region ($r < w$). First, if the particle is located in the central part of the laser beam ($r\ll w$), the restoring force and torque are very weak, thereby allowing  it to propel itself radially outwards in a rather directed manner where ${\bf{n}}$ is quasi-parallel to ${\bf{r}}$, \emph{i.e.} $\varphi \approx \pi$. This leads to a systematic increase in its radial distance, which translates into a more and more unstable situation for ${\bf{n}}$ because rotational diffusion can easily produce a perturbation around $\varphi = \pi$ either clockwise or counterclockwise. When this happens, the optical torque tends to quickly reorient ${\bf{n}}$ antiparallel to ${\bf{r}}$, which in combination with the restoring optical force facilitates the self-propulsion of the particle toward the center of the beam. As $r$ decreases, the effect of optical force and torque also decreases and the particle moves actively in a directed manner, thus repeating the process over and over again. 

\begin{figure}[h]
\centering
\includegraphics[width=0.8\columnwidth]{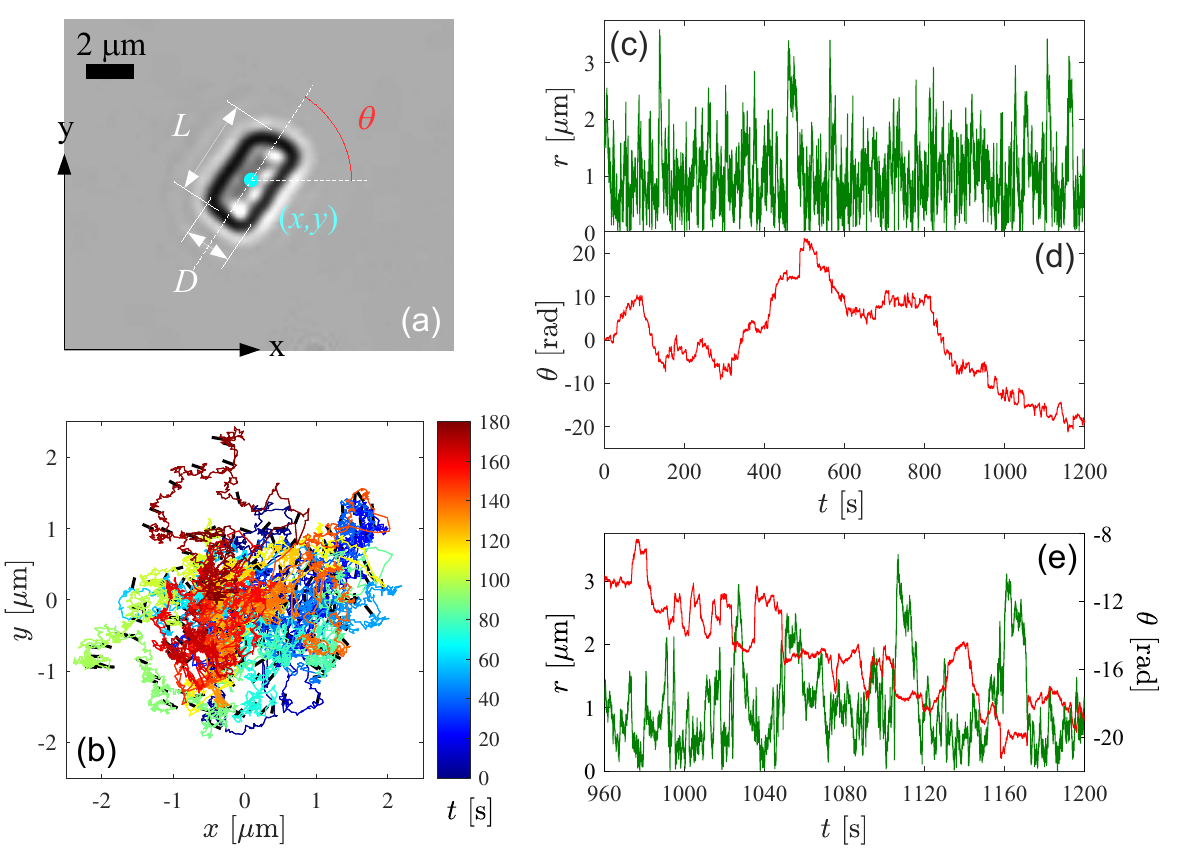}
  \caption{(a) Picture of a colloidal rod of length $L = 4\,\mu$m and diameter $D = 2\,\mu$m half-coated with a $ 100$~nm tick carbon cap (dark side on the left). The coordinates $(x,y)$ and $\theta$ used to characterize its motion in the converging laser beam are also depicted. (b) Example of a trajectory segment of 3 minutes of a Janus rod self-propelling in the beam at $P = 1$~mW. The orientation of the projection of its main axis onto the $xy$ plane defined by the azimuthal angle $\theta$ is traced as bars every 0.5~s. The colormap depicts the time elapsed during the interval $0 \le t \le 180$~s. [(c)-(d)] Stochastic time evolution over $0\le t \le 1200$~s in the presence of the laser beam at $P = 1$~mW of: (c) the radial position $r(t)$ of the same Janus rod; (d) its azimuthal angle $\theta(t)$. (e) An expanded view of the coordinates $r(t)$ (green) and $\theta(r)$ (red) plotted in (c) and (d), respectively, over the last 240~s.}
  \label{fig:4}
\end{figure}

\subsection{Active confinement of rod-shaped Janus particles}\label{subsec:anisop}

Finally, we demonstrate that an active trapping mechanism akin to that previously described for spherical Janus beads arises in the case of more anisotropic Janus colloidal particles, such as rods that are half-coated with carbon. To produce them, we follow the same procedure as described in section \ref{sect:exp}, which results in a partial carbon coating along the main axis of the rods, as illustrated in figure \ref{fig:4}(a). The carbon-coated rods are also dispersed in ultrapure water and exposed to the same converging laser beam as the one sketched in figure \ref{fig:1}(a) at powers at which they remain near the lower surface of the sample cell. We observe that in the presence of the laser, the rods can rotate in three dimensions, and then the detection of their instantaneous propulsion orientation, defined as the unitary vector pointing from the capped to the uncapped side like in the case of Janus beads, is generally challenging. Hence, to characterize their active motion, we only track their center of mass $(x,y)$ and the azimuthal angle $\theta$ between the projection of their main axis onto the $xy$ plane and the $x$ axis, which are outlined in figure \ref{fig:4}(a). 

Similarly to the behavior of the Janus beads in the converging beam, the carbon-capped rods can be stably trapped owing to the interplay between self-propulsion and optical forces and torques above a threshold laser power that depends on the rod size. This is exemplified in figure \ref{fig:4}(b), which shows a portion over 3~minutes of the trajectory $(x,y)$ of a rod of length $L = 4\,\mu$m and diameter $D = 2\,\mu$m. This active trajectory qualitatively resembles the self-propelled motion of Janus beads in the confining optical potential, like the trajectory displayed in figure \ref{fig:3}(c), namely, rather directed paths intertwined with looping ones occurring when the particle moves away from the beam center. We also depict some instantaneous orientations of the main axis of the rod determined by the angle $\theta$, which reveal that the particle undergoes large orientational changes that are not simply caused by rotational diffusion, but are the result of the optical torques acting on it.
Examples of stochastic changes over time $t$ of the particle's radial distance $r(t) = \sqrt{x(t)^2 + y(t)^2}$ and angle $\theta(t)$,
are plotted in figures \ref{fig:4}(c) and \ref{fig:4}(d), respectively. We find that the rods can propel themselves radially away from the central part of the region close to the glass surface that is illuminated by the laser, but they eventually
return without fully escaping it due to the restoring effect of the converging beam. The expanded view of the radial and angular trajectories in figure \ref{fig:4}(e) reveals that the resulting orientational changes are not as abrupt as the reversals of approximately $\pm \pi$~rad shown in figures (\ref{fig:2})(c) and (\ref{fig:2})(f) for Janus beads. The origin of this difference is that the out-of-plane rotation of the main axis of the Janus rods under the action of the laser randomizes the orientation of their carbon cap, thus disturbing the alignment of their propulsion direction towards the center of the beam as occurs with the Janus spheres. This in turn hinders the periodicity of the rotational dynamics of Janus rods in comparison to the case of carbon-coated beads, thus preventing the full development of oscillatory motion. It is interesting to note that, despite the additional anisotropy in the dynamics induced by the particle shape, the combination of thermophoretic self-propulsion and restoring optical forces and torques also leads to active back and forth motion of the colloidal carbon-capped rods, thus enabling their stable dynamical trapping within the converging beam. 
 
\section{Conclusions}

In this paper, we have experimentally investigated the self-propelled motion of dielectric  colloidal particles half-coated with a thin layer of carbon in a converging laser beam. We have found that, unlike metal-capped Janus particles that have been utilized in numerous experimental investigations reported in the literature, the carbon-coated particles studied here exhibit a distinctive active motion that allows them to explore in a back and forth manner their optical potential landscape. As a first example, spherical Janus beads exhibit a quasi-two-dimensional motion that gives rise to oscillations within the optical confinement, whose frequency increases monotonically with the propulsion speed. The analysis of their mean-square displacement reveals the existence of four regimes of their translational motion depending on the timescale, namely thermal diffusion, ballistic motion, oscillatory behavior, and confinement.
The experimental results can be described by a phenomenological stochastic model that includes the effect of a torque that tends to align the particle orientation toward the center of the laser beam and whose magnitude increases with the radial distance. As a second example, carbon-coated rods exhibit a qualitatively similar dynamical trapping within the converging laser beam, even though the periodicity of their motion is hindered by the three-dimensional rotation in the presence of the laser. 

In recent years there has been a rising theoretical and experimental interest in studying the self-propelled motion of Janus colloidal particles in diverse energy landscapes, in particular in confining potentials, as it represents a minimal setting of active matter systems that displays nontrivial effects due to their non-equilibrium nature. The results presented here illustrate the emergence of striking dynamical behavior in a trapping potential, such as self-sustained oscillations, which originates from the interplay between self-propulsion and restoring optical forces and torques. Therefore, these effects, which are not generally considered in simpler theoretical models, could be of interest from a fundamental viewpoint as well as for practical applications, e.g. in Brownian heat engines with active particles as working substance~\cite{speck2022,nalupurackal2023,cotevalencia2025}. Other physical aspects of this active system, such as its motion in
viscoelastic baths \cite{darabi2023,ginot2025} or across multistable potentials \cite{ferrer2024,ferrer_2024}, are also relevant to the field of active matter and will be the subject of further experimental and theoretical research.

\ack{We thank Diego Quiterio-Vargas and Arturo Rodr\'iguez-G\'omez for their assistance in production of Janus particles.}

\funding{We acknowledge the support from DGAPA-UNAM PAPIIT Grant No. IN110324 and PIIF-2-24.}

\roles{F Darabi: conceptualization, data curation, formal analysis, investigation, methodology, software, validation, visualization. J R Gomez-Solano: conceptualization, data curation, formal analysis, funding acquisition, investigation, methodology, project administration, resources, software, supervision, validation, visualization, writing—original draft.}

\data{
The data that support the findings of this study are available upon
reasonable request from the authors.}

\appendix

\section{Velocity autocorrelation function}\label{app:1}

\renewcommand{\thefigure}{A\arabic{figure}}

\setcounter{figure}{0}

\begin{figure}[h]
\centering
\includegraphics[width=0.9\columnwidth]{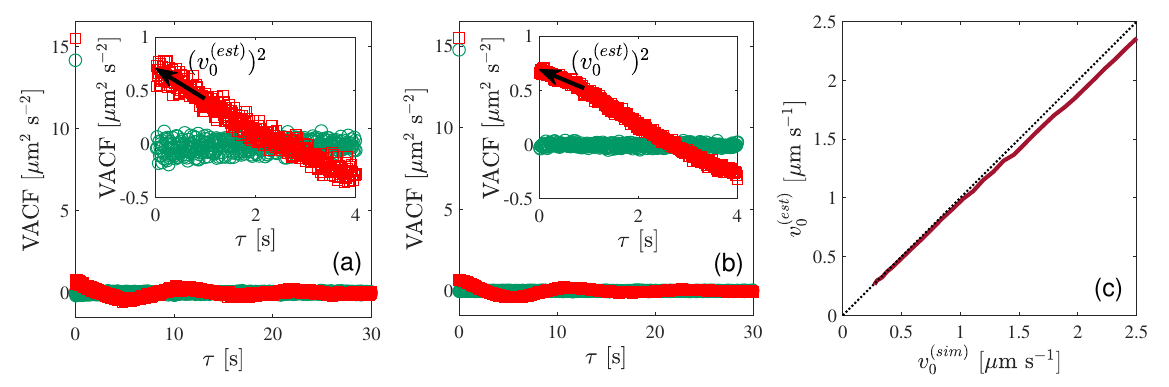}
  \caption{(a)  Experimental velocity autocorrelation function of a Janus bead recorded at $f_s = 100$~Hz freely moving at  $P = 0$ ($\circ$) and in the converging laser beam at $P = 0.5$~mW ($\Box$). Inset: expanded view of the main figure that highlights the extrapolation $(v_0^{(est)})^2$ to $\tau = 0$ of the oscillatory part of the VACF to estimate $v_0$ and $D_T$. (b) Numerical velocity autocorrelation function obtained from the simulation of equations (\ref{eq:xy}) and (\ref{eq:theta}) for a particle freely diffusing with $v_0^{(sim)} = 0$ ($\circ$) and actively moving under confinement at $v_0^{(sim)} = 0.840\,\mu$m~s$^{-1}$ ($\Box$). Inset: expanded view of the main figure that highlights the extrapolation $(v_0^{(est)})^2$ to $\tau = 0$ of the oscillatory part of the VACF. (c) Comparison between the values $v_0^{(est)}$ of the characteristic propulsion speed estimated through the VACF and those used in the simulations of the model, $v_0^{(sim)}$ (solid line). The identity function (dotted line) depicts perfect agreement.} \label{figure:A}
\end{figure}

In this Appendix, we provide some further details on the calculation of the characteristic propulsion speed $v_0$ of the confined particles and their short-time diffusion coefficient from their velocity autocorrelation function (VACF) calculated at two times, $t$ and $t + \tau$, which is defined as $\langle \dot{{\bf{r}}}(t +\tau) \cdot \dot{{\bf{r}}}(t) \rangle_t$, with $\dot{{\bf{r}}}$ the instantaneous particle velocity, $\tau \ge 0$ the duration of the interval between the two times, and $\langle \ldots \rangle_t$ a time average over the initial time $t$.
Due to the diffusive nature of the overdamped motion of the Janus colloids in water in the absence of activity induced by the laser, the VACF consists of a single Dirac-delta peak at $\tau = 0$, i.e. $\langle \dot{{\bf{r}}}(t +\tau) \cdot \dot{{\bf{r}}}(t) \rangle_t = 4D_T \delta(\tau)$. At a finite sampling frequency $f_s$, the instantaneous velocity can be estimated from the finite time difference between consecutive positions ${\bf{r}}$ as $\dot{{\bf{r}}}(t) = f_s \left[{\bf{r}}(t + f_s^{-1}) - {\bf{r}}(t) \right]$, and then the corresponding VACF has a finite peak of height $4D_T f_s$ at $\tau=0$, as shown in figure \ref{figure:A}(a) for the experimental VACF of a freely moving Janus bead ($P = 0$). Moreover, when the laser induces a persistent velocity in the particle motion with a variance $v_0^2$, this is added to the finite peak at $\tau=0$. In the case of the oscillatory dynamics of the Janus particles studied here, this also leads to oscillations of the VACF. Therefore, the characteristic propulsion speed $v_0$ can be estimated from the square root of the linear extrapolation $(v_0^{(est)})^2$ to $\tau = 0$ of the VACF at small values of $\tau > 0$, as experimentally illustrated in figure \ref{fig:4}(a) for a Janus particle moving under the action of the laser beam at $P = 0.5$~mW. In addition, the corresponding translational diffusion coefficient in the presence of activity can be estimated from the height of the peak at $\tau = 0$ as $D_T = \frac{1}{4f_s} \left[ \langle |\dot{{\bf{r}}}(t)|^2 \rangle_t - (v_0^{(est)})^2 \right]$.

We verified this method by means of the calculation of the VACF for numerically simulated trajectories that evolve stochastically in time according to equations (\ref{eq:xy}) and (\ref{eq:theta}). As shown in figure \ref{figure:A}(b), these numerical VACF curves reproduce the behavior of the experimental ones with a Dirac-delta peak at $\tau = 0$ followed by damped oscillations when $v_0 >0$. Furthermore, figure \ref{figure:A}(c) shows that the values $v_0^{(est)}$ of the characteristic propulsion speed estimated from the extrapolation of the VACF are slightly smaller but very close to those set in the simulation, $v_0^{(sim)}$, thus providing a reliable method to estimate $v_0$ in the case of the experimental data. 

\setcounter{figure}{0}



\end{document}